# Teaching Quantum Formalism and Postulates to First-Year Undergraduates

Jeremy Levy and Chandralekha Singh
*Department of Physics and Astronomy, University of Pittsburgh, Pittsburgh, PA 15260 USA.*
(Dated: December 2, 2024)

Traditional approaches to undergraduate-level quantum mechanics require extensive mathematical preparation, preventing most students from enrolling in a quantum mechanics course until the third year of a physics major. Here we describe an approach to teaching quantum formalism and postulates that can be used with first-year undergraduate students and even high school students. The only pre-requisite is a familiarity with vector dot products. This approach enables students to learn Dirac notation and core postulates of quantum mechanics at a much earlier stage in their academic career, which can help students prepare for careers in quantum science and engineering and advance the Second Quantum Revolution.

## INTRODUCTION

Quantum mechanics is arguably the most important course that an undergraduate physics major will take. It is central to understanding many aspects of physics, including astrophysics, particle physics, solid-state and materials science, photonics, and, more recently, quantum information science, engineering, and technology.

There is enormous enthusiasm for lowering the entry barriers for quantum science and engineering, in order to prepare a new generation of students to address the challenges of the Second Quantum Revolution [1, 2]. Many educational activities are aimed at younger and less-technically-advanced students, with the hope of preparing them for future careers in quantum industries.

In most university curricula, quantum mechanics is taught as a one-semester or two-semester course, prefaced by modern physics and a full year of introductory physics. In preparation for learning quantum mechanics, students must first take courses in advanced calculus, complex analysis, differential equations, and linear algebra.

Why do we wait so long to teach quantum mechanics? Apart from curricular inertia, there is a broad consensus that students simply cannot learn quantum mechanics at an earlier stage of their academic career. In the early days of quantum theory (i.e., the dawn of the First Quantum Revolution), quantum mechanics was not taught to undergraduates, nor was it offered as a core graduate course[3]. One of the reasons why quantum education was delayed until the advanced graduate level was that the mathematical notation at the time was not well developed. Dirac provided a pedagogical service to the physics community by creating his namesake "Dirac notation" [4], which helped to codify and simplify operations that are often obscured by representation-specific approaches or matrix-only representations of the Time-Dependent Schrödinger Equation (TDSE). However, despite this improved notation, a century after the introduction of quantum mechanics, it is still a highly challenging course for students.

Many introductory textbooks adopt a "spins-first" approach to quantum mechanics to help students appreciate the mathematical structure of quantum mechanics[5, 6], while setting the stage for learning about quantum information. Two-state quantum systems are not just spins, but also "qubits" which can be used to store quantum information. Dirac notation is generally not introduced in modern physics courses that cover quantum mechanics; introductory quantum textbooks have tended to use Dirac notation sparingly, although some more modern textbooks do employ Dirac notation more thoroughly [7, 8].

Efforts to make quantum mechanics more accessible to students, so that they can be introduced to the subject at an earlier state, have been described by, for example, Zollman et al. [9]. Rudolf has developed an accessible guide to quantum concepts (*Q is for Quantum*, Ref. [10]) which are necessary for quantum computing. Hands-on approaches often involve visualization of wavefunctions and de-emphasize the formal aspects of learning quantum mechanics. Other efforts to introduce quantum mechanics include the development of a variety of quantum-themed games [11–24] designed to teach various aspects of quantum mechanics.

Here we describe an approach to teaching quantum mechanics to students who have no more than one semester of introductory physics. The material was developed as a one-credit course that gives students an introduction to how quantum mechanics works and provides a sound foundation for future studies. We create a bridge to learning Dirac formalism by extending the vector notation that students learn in an introductory calculus-based physics course. This extension enables a one-to-one map with Dirac notation, including the representation of operators. Elements of quantum theory are illustrated using a "Bloch cube", which is a simplified version of the Bloch sphere. The Bloch cube enables introductory students to manipulate quantum states with their hands, thus building intuition about many advanced quantum concepts. This approach was developed and refined over a three-year period with three groups of 5-10 first-year undergraduate students by J.L. in collaboration with C.S. Students have consistently provided very positive feedback in post-instruction surveys regarding this approach.

## BREAKING DOWN THE SCHRÖDINGER EQUATION

It is helpful to break down the TDSE (Eq. 1) into five components, which correspond roughly to five mathematical pre-requisites, summarized in Table I.

$$i\hbar \frac{\partial |\psi\rangle}{\partial t} = \hat{H} |\psi\rangle \quad (1)$$

TABLE I. Breaking down the TDSE.

| TDSE Component | Mathematical Requirements |
|---|---|
| $i$ | Complex numbers |
| $\hbar$ | Algebra |
| $\partial/\partial t$ | Calculus, differential equations |
| $|\psi\rangle$ | Vectors |
| $\hat{H}$ | Linear algebra |

$i$: Complex arithmetic, multiplication, and complex conjugation are taught in calculus-based or honors introductory physics courses.

$\hbar$: The Planck constant is just a number (with units), emblematic of symbolic representation of numbers, and ordinary algebraic manipulations.

$\partial/\partial t$: Calculus is often a co-requisite in introductory physics courses. At a more advanced stage, students learn about coupled and partial differential equations.

$|\psi\rangle$: Vectors are used extensively in introductory physics to describe the position and dynamics of objects, and other fields (e.g., magnetic, electric) in three spatial dimensions. Non-calculus-based courses often skip the concept of unit vectors, whereas in calculus-based curricula they are well developed and used.

$\hat{H}$: Operators, linear algebra and matrix manipulation are usually avoided in introductory physics courses.

To introduce quantum mechanics to introductory physics students, we need to choose what is essential and what can be left for more advanced instruction. Of the mathematics skill requirements just outlined, the skills least likely to be possessed by introductory students are linear algebra and differential equations.

The linear algebra required to use Dirac notation can be introduced by making an explicit analogy with unit vectors. This analogy can be used not only to teach about inner products, but also outer products and identity and rotation operators, as described in Sec. III-IV. The Bloch cube is introduced in Sec. V to provide a tangible representation of quantum, states and the Born rule.

The requirement for differential equations is eliminated because the TDSE can be "integrated out" to yield unitary operators $\hat{U} = \exp[-i\hat{H}t/\hbar]$ with the property that

$$\hat{U} |\psi(0)\rangle = |\psi(t)\rangle \quad (2)$$

assuming that $\hat{H}$ is time-independent. Eq. 2 serves as a substitute for Eq. 1, which students do not need to learn how to solve until later in their academic career. While understanding the TDSE is essential to understanding the physical mechanisms by which states are transformed, students at earlier stages in their physics education can learn to work with the operators (in this work, rotation operators) that represent these transformations. Sec. VI shows how the rotation operator is introduced and illustrated with the Bloch cube. Then in Sec. VII-VIII, we outline the lecture sequence for this one-credit course and show how it compares to traditional instruction.

## VECTORS AND DOT-VECTORS

Introductory students learn about unit vectors and scalar products, which are used to convert vector equations into a set of scalar equations. Here we review and then extend the notation to make it more compatible with Dirac "bra-ket" notation. We will restrict ourselves to two dimensions, which is sufficient for mapping onto two-state quantum systems. Unit vectors obey the standard set of orthogonality relations: $\hat{x} \cdot \hat{x} = 1$, $\hat{x} \cdot \hat{y} = 0$, $\hat{y} \cdot \hat{x} = 0$, $\hat{y} \cdot \hat{y} = 1$. A general vector $\vec{V}$ can be written as:

$$\vec{V} = V_x \hat{x} + V_y \hat{y} \quad (3)$$

where $V_x = \hat{x} \cdot \vec{V}$ and $V_y = \hat{y} \cdot \vec{V}$. Complete sets of orthonormal unit vectors define coordinate systems.

To help students understand bra-ket notation, we start by introducing a new type of vector, called the "dot-vector", in which the "·" is attached to the unit vectors to define a new vector space. The dot-vector can be disarmingly introduced by making reference to the Dr. Seuss story about Sneetches [25]. In that children's story, there are two kinds of Sneetches: Star-Bellied Sneetches and Plain-Bellied Sneetches. Other than the fact that one sports a star on its belly, they look identical. By analogy, we can define, for every unit vector $\hat{x}$ or $\hat{y}$, a corresponding dot-vector $\hat{x}\cdot$ and $\hat{y}\cdot$, as shown in Figure 1. A general dot-vector $\vec{W}\cdot$ in the two-dimensional dot-vector space can be written as:

$$\vec{W}\cdot = W_x \hat{x}\cdot + W_y \hat{y}\cdot \quad (4)$$

where $W_x$ and $W_y$ are scalar coefficients. The scalar product between $\vec{W}\cdot$ and $\vec{V}$ is given by:

$$\vec{W} \cdot \vec{V} = (W_x \hat{x}\cdot + W_y \hat{y}\cdot)(V_x \hat{x} + V_y \hat{y}) = W_x V_x + W_y V_y \quad (5)$$

where we have taken advantage of the orthonormality relations defined above. The class of operations that students learn to manipulate vectors applies similarly to dot-vectors. In this sense, the distinction between vectors and dot-vectors is superficial, not unlike the stars that some Sneetches wear on their bellies.

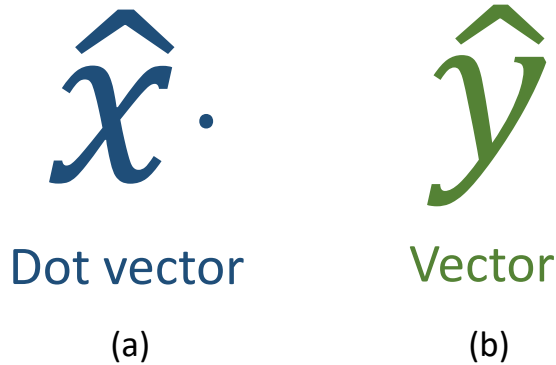

FIG. 1. Dot-Vec notation developed to help introduce Dirac notation. (a) Dot-vector $\hat{x}\cdot$. (b) vector $\hat{y}$. Products of dot-vectors and vectors can yield scalars or operators, depending on their relative ordering.

By combining vectors and dot-vectors in the opposite order, we can create operators which transform vectors into other vectors. An extremely useful operator is the identity operator, which can be written as follows:

$$\mathbf{1} = \hat{x}\hat{x}\cdot + \hat{y}\hat{y}\cdot. \quad (6)$$

Acting the identity operator $\mathbf{1}$ on the vector $\vec{V}$ yields:

$$(\hat{x}\hat{x}\cdot + \hat{y}\hat{y}\cdot)\vec{V} = \hat{x}\hat{x}\cdot\vec{V} + \hat{y}\hat{y}\cdot\vec{V} = V_x\hat{x} + V_y\hat{y}. \quad (7)$$

where $V_x = \hat{x}\cdot\vec{V}$ and $V_y = \hat{y}\cdot\vec{V}$. The identity operator provides an explicit method for resolving a vector into components, which is something that introductory students are often asked to do when solving vector equations.

Another type of operator is one that produces a rotation in the $xy$ plane. For example, the operator $\mathbf{R}$ rotates vectors counterclockwise by 90°.

$$\mathbf{R} = \hat{y}\hat{x}\cdot - \hat{x}\hat{y}\cdot. \quad (8)$$

The rotation properties can be seen by its effect on three test vectors, illustrated in Fig. 2.

An important conceptual leap which is necessary for understanding quantum mechanics is the mapping between orthogonal vectors and distinct states of a system. Unit vectors in quantum mechanics can represent mutually distinguishable states and not just directions in space. For example, we can imagine that $\hat{x}$ represents the "heads" state of a coin, while $\hat{y}$ represents the "tails" state.

The action of flipping over a coin can be described by the operator $\mathbf{F} = \hat{x}\hat{y}\cdot + \hat{y}\hat{x}\cdot$.

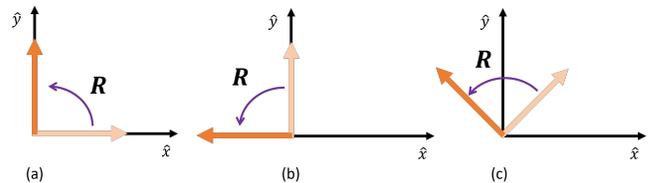

FIG. 2. Effect of $\mathbf{R}$ operator on (a) $\hat{x}$, (b) $\hat{y}$, (c) $(\hat{x}+\hat{y})/\sqrt{2}$.

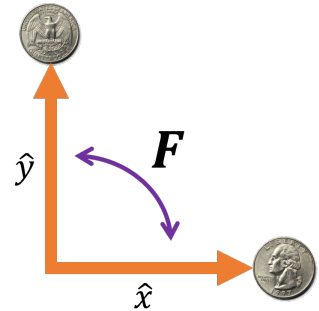

FIG. 3. Unit vectors can represent states of matter. Operator $\mathbf{F}$ flips the state of a coin.

## DIRAC NOTATION

After extending the existing vector notation to include dot-vectors (which we will refer to as "Dot-Vec" notation), the analogy with Dirac (bra-ket) notation can be introduced. Ket states are associated with unit vectors, and Bra states are associated with unit dot-vectors. The scalar products defined for $\hat{x}$ and $\hat{y}$ are mapped onto the scalar product relations for Dirac bras and kets. The analogy is summarized in Table II.

### Operators

With this identification, we can use some of the tools developed for vectors to resolve a general state $|\psi\rangle$ into components along $|+z\rangle$ and $|-z\rangle$ using the identity operator $\hat{1} = |+z\rangle\langle+z| + |-z\rangle\langle-z|$.

TABLE II. Dot-Vec / Bra-Ket notation analogy.

| Dot-Vec | Bra-Ket |
|---|---|
| $\hat{x}$ | $|+z\rangle$ |
| $\hat{y}$ | $|-z\rangle$ |
| $\hat{x}\cdot\hat{x} = 1$ | $\langle +z|+z\rangle = 1$ |
| $\hat{x}\cdot\hat{y} = 0$ | $\langle +z|-z\rangle = 0$ |
| $\hat{y}\cdot\hat{x} = 0$ | $\langle -z|+z\rangle = 0$ |
| $\hat{y}\cdot\hat{y} = 1$ | $\langle -z|-z\rangle = 1$ |
| $\hat{x}\hat{x}\cdot$ | $|+z\rangle\langle +z|$ |
| $\hat{x}\hat{y}\cdot$ | $|+z\rangle\langle -z|$ |
| $\hat{y}\hat{x}\cdot$ | $|-z\rangle\langle +z|$ |
| $\hat{y}\hat{y}\cdot$ | $|-z\rangle\langle -z|$ |

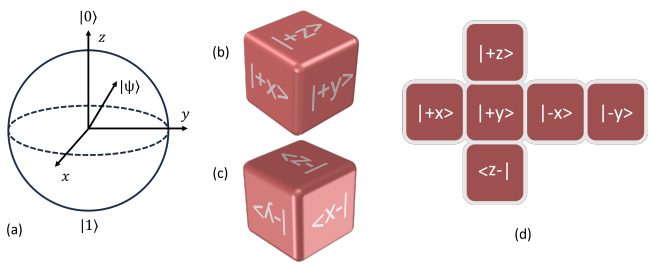

FIG. 4. Bloch sphere and Bloch cube. (a) Bloch sphere. (b) Top view of Bloch cube. (c) Bottom view of Bloch cube. (d) Fold-out pattern of Bloch cube.

$$\hat{1}|\psi\rangle = |+z\rangle\langle+z|\psi\rangle + |-z\rangle\langle-z|\psi\rangle = \psi_+|+z\rangle + \psi_-|-z\rangle \tag{9}$$

where $\psi_+ = \langle+z|\psi\rangle$ and $\psi_- = \langle-z|\psi\rangle$.

When students are comfortable with scalar products and operators using Dirac notation, it is possible to discuss how it relates to quantum physics. In keeping with a "spins-first" approach, we restrict our attention to the two-dimensional space spanned by $|+z\rangle$ and $|-z\rangle$, which can also labeled $|0\rangle$ and $|1\rangle$ in the context of quantum information and qubits, to describe a physical system that can be in one of two possible distinct states. We avoid describing any specific physical system, e.g., spin-1/2, since these concepts can increase the cognitive load of students. Additionally, in the first semester of college-level introductory physics, students have not been exposed to the idea of magnetic fields or spins. The abstract nature of the labels used to describe the two-state system is compensated by a "hands-on" approach involving a Bloch Cube, which is described in the next Section below.

## BLOCH CUBE

Quantum states in two dimensions are often represented in terms of the Bloch sphere (Figure 4(a)). Here we simplify the Bloch sphere by fashioning it into a six-sided "Bloch cube" (Figure 4(b,c)). Opposite faces of the cube represent distinct states, and are labeled $|\pm n\rangle$, $n = x, y, z$. The allowed states are restricted to the six possible states represented on the Bloch cube. While only certain states are represented, the set is sufficiently rich to illustrate most of the important quantum concepts. Bloch cubes (Figure 5), like a deck of cards, can be used in multiple exercises or games to illustrate important quantum concepts.

It is important for students to understand that, even though we work with states labeled $|+x\rangle$, $|+y\rangle$, and $|+z\rangle$, none of these states are "orthogonal" to each other. Students should be told over and over that distinct states are those which are situated on opposite faces of the Bloch Cube. There are many opportunities for instructors to emphasize this crucial point during instruction, but the Bloch Cube can help remind students of this.

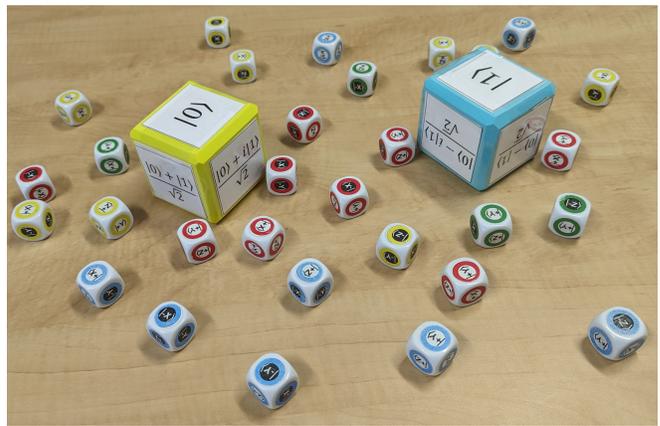

FIG. 5. Bloch Cubes. The smaller ones are made from blank ceramic cubes and covered with laser-printed labels. The larger Bloch Cubes use a different "qubit" labeling in which $|+z\rangle = |0\rangle$ and $|-z\rangle = |1\rangle$, and other states are expressed in this basis.

The Bloch Cube can be used to represent the quantum state of two-state system. We will work with a convention that the state of a Bloch Cube is whatever side is facing upwards. This convention is useful for students who are working with the Bloch Cubes themselves. The convention can be adjusted if one is giving instruction to a class, in which the side of the Bloch Cube that is facing the students can serve as the Bloch Cube state.

## Quantum Measurements and the Born Rule

The Born rule predicts the likelihood of a measurement outcome, and it can be presented as a postulate. If the system is in a state $|\psi\rangle$, the probability of measuring it to be $|\phi\rangle$ is given by

$$P(|\psi\rangle \to |\phi\rangle) = |\langle\phi|\psi\rangle|^2 \tag{10}$$

To illustrate this idea with the Bloch cube, we can define (according to standard conventions) the following superposition states (Eq. 11):

$$|+x\rangle = \frac{1}{\sqrt{2}}(|+z\rangle + |-z\rangle) \tag{11a}$$

$$|-x\rangle = \frac{1}{\sqrt{2}}(|+z\rangle - |-z\rangle) \tag{11b}$$

$$|+y\rangle = \frac{1}{\sqrt{2}}(|+z\rangle + i|-z\rangle) \tag{11c}$$

$$|-y\rangle = \frac{1}{\sqrt{2}}(|+z\rangle - i|-z\rangle) \tag{11d}$$

TABLE III. Born rule probabilities $|\langle \Box | \Box \rangle|^2$ for Bloch cube states.

| Question | Final State | $|+z\rangle$ | $|-z\rangle$ | $|+x\rangle$ | $|-x\rangle$ | $|+y\rangle$ | $|-y\rangle$ |
|---|---|---|---|---|---|---|---|
| | | \multicolumn{6}{c}{Initial State} | | | | | |
| Z | $\langle +z|$ | 100% | 0% | 50% | 50% | 50% | 50% |
| Z | $\langle -z|$ | 0% | 100% | 50% | 50% | 50% | 50% |
| X | $\langle +x|$ | 50% | 50% | 100% | 0% | 50% | 50% |
| X | $\langle -x|$ | 50% | 50% | 0% | 100% | 50% | 50% |
| Y | $\langle +y|$ | 50% | 50% | 50% | 50% | 100% | 0% |
| Y | $\langle -y|$ | 50% | 50% | 50% | 50% | 0% | 100% |

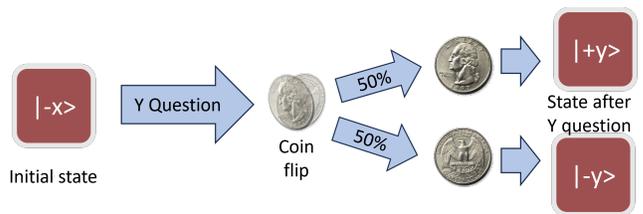

FIG. 6. Example illustrating measurement using the Bloch cube. The initial state is $|-x\rangle$, and the Y Question is asked. A coin is flipped and if it lands heads, the Bloch cube is oriented to the $|+y\rangle$ state; if it lands tails, the Bloch cube is oriented in the $|-y\rangle$ state.

If a system is in the $|+x\rangle$ state, students can calculate that the probability of measuring $+z$ is given by

$$P(|+x\rangle \to |+z\rangle) = |\langle +z|+x\rangle|^2 = \frac{1}{2} \qquad (12)$$

The full set of probabilities for measurement is summarized in Table III. The rationale behind Eq. 11 is not provided to students by default, but students can be asked to work out the entries in Table III based on those definitions, and can be asked to think about alternative definitions that would be consistent with Table III.

To simulate the measurement process with a Bloch cube, students first choose an initial state, which by convention can be the face-up state of the Bloch cube after rolling. One of three possible questions can be asked (Eq. 13(a-c)).

$$\text{X Question:} |+x\rangle \text{ state or } |-x\rangle \text{ state?} \qquad (13\text{a})$$
$$\text{Y Question:} |+y\rangle \text{ state or } |-y\rangle \text{ state?} \qquad (13\text{b})$$
$$\text{Z Question:} |+z\rangle \text{ state or } |-z\rangle \text{ state?} \qquad (13\text{c})$$

If the question matches the state (e.g., Z question for $|+z\rangle$ state), then the state is unchanged. If the question type differs from the face-up state (e.g., X question for $|+z\rangle$ state), the Bloch Cube is rolled and the state changes to $|+x\rangle$ if the roll lands on a positive face; otherwise, the state changes to $|-x\rangle$. A flowchart is illustrated in Figure 6. An alternative to using a Bloch Cube to generate a probabilistic outcome would be to flip a coin.

## QUANTUM DYNAMICS

Solving the TDSE, even with a two-state quantum system, generally involves calculus, differential equations, and matrices, which are beyond the reach of most students taking introductory physics. However, the TDSE can be integrated to yield unitary operators that represent the formal solutions of the TDSE and correspond to physical rotations. We represent unitary operators as $\hat{U}$, where the "hat" is not to be confused with the unit vectors and dot-vectors from before. A unitary operator will in general take the form $\hat{U} = e^{i\phi_+}|+u\rangle\langle+z| + e^{i\phi_-}|-u\rangle\langle-z|$, so that $\hat{U}|+z\rangle = e^{i\phi_+}|+u\rangle$ and $\hat{U}|-z\rangle = e^{i\phi_-}|-u\rangle$, where $\{|+u\rangle, |-u\rangle\}$ form an orthonormal basis. However, if the initial and final states are restricted to the six faces of the Bloch cubes, this constrains the rotation angles to be integer multiples of 90°. Using this restriction, there is no need to teach students about the TDSE or to define unitary operators. The resulting system is simple enough to be readily understood and yet rich enough to illustrate how operators transform states.

We begin by defining rotation operators that rotate the Bloch cube about the $X$, $Y$, and $Z$ axes by 90° in the clockwise direction (Figure 7). We can represent the $\hat{Z}$ unitary operation corresponding to 90° rotation about the $Z$ axis in two equivalent ways:

$$\hat{Z} = |+z\rangle\langle+z| + i|-z\rangle\langle-z| \qquad (14\text{a})$$
$$\hat{Z} = -i|+z\rangle\langle+z| + |-z\rangle\langle-z| \qquad (14\text{b})$$

The two forms of $\hat{Z}$ differ only by a constant multiplicative factor, which we explain is an arbitrary mathematical choice with no physical effects, similar to the choice of zero potential energy in classical mechanics problems. Note that with these definitions, $\hat{Z}^4 = +1$; this is the rotation operator for a Bloch cube and not for a spin-1/2 state. Acting with $\hat{Z}$ or its inverse $\hat{Z}^{-1}$ on $|+z\rangle$ and $|-z\rangle$, does not change the state except for a factor of $\pm i$. The fact that $|+z\rangle$ and $|-z\rangle$ are eigenstates of $\hat{Z}$ is intuitively seen because the Bloch cube is being rotated about the $Z$ axis and hence does not transform into another state.

To understand the effect of $\hat{Z}$ on the other four Bloch cube states, one can rotate the Bloch cube 90° clockwise about the $Z$ axis. In general, $\hat{Z}$ permutes the states as follows: $|+x\rangle \to |+y\rangle \to |-x\rangle \to |-y\rangle \to |+x\rangle$. Students can see this mathematically by rewriting the $\hat{Z}$ operator in terms of its action on, for example, the $|+x\rangle$ and $|-x\rangle$ basis states. Students can also show mathematically that

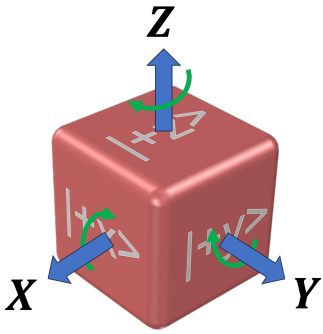

FIG. 7. Illustration of Bloch cube clockwise 90° rotations about $X$, $Y$, and $Z$ axes.

$\hat{Z}$ can be expressed as follows:

$$\hat{Z} = |+y\rangle\langle+x| + |-y\rangle\langle-x| \qquad (15a)$$

$$\hat{Z} = |-x\rangle\langle+y| + |+x\rangle\langle-y| \qquad (15b)$$

Expressions like these are readily understood by examining the Bloch cube and remembering the definition for the $\hat{Z}$ unitary rotation. Similar expressions can be found for rotations about the $x$ and $y$ directions.

In general, the Bloch cube serves as a "quantum abacus" for computing a series of unitary operations. Students can discover, for example, the non-commutative nature of these operations by writing down a sequence of operations in various orders and showing that the Bloch cube ends up in a different state depending on the order of operations.

## LECTURE SEQUENCE

This approach to teaching quantum mechanics was used in a one-credit course, but the level of detail can be expanded or compressed depending on the amount of time available. The development of this course took place with students who participated in a program titled "First Experiences in Quantum" (FEQ), held in the Spring semester at the University of Pittsburgh. Most students were first-year undergraduates who were also signed up to work on a project in the research group of a willing faculty member for 1-2 course credits. Because any grade was assigned by the research advisor, the only course assessment was a midterm test; there was no final exam and no out-of-class assignemnts. Below is a list of topics covered in the lectures.

1. Introduction and overview
2. Vectors and unit vectors in two dimensions
3. Dot products and introduction to Dot-Vec notation
4. More examples of Dot-Vec notation and coordinate transformations
5. Dirac notation and analogy with Dot-Vec notation
6. Bloch cube and representation of quantum states
7. Bloch cube and Born rule
8. Bloch cube and quantum dynamics
9. Proving that unitary operators are represented by Bloch cube rotation
10. Final assessment

Additional suggestions about how instructors may follow the approach described here are given in the Supplementary Material. Additional topics that cover more advanced topics can be taught using Bloch Cubes, including density matrix formalism, electron spin resonance, quantum key distribution, and elements of quantum field theory. These topics require the use of multiple Bloch Cubes. For example, the density matrix is represented by a set of Bloch cubes that represent the possible states and their respective probabilities. The completely mixed state would have half of the Bloch Cubes in a $|+z\rangle$ state and half in the $|-z\rangle$ state. Quantum key distribution (BBM92 type [26]) would use Bloch Cubes to represent photon states measured Alice and Bob, as well as the outcome of joint measurements. Electron spin resonance involves rotations of states, which can be simulated using Bloch Cubes. Inhomogeneous dephasing can be represented "digitally" by allowing some states to rotate faster than others and advance by one quarter turn. A "pi pulse" applied to all the spins followed by more spinning will cause refocusing and spin-echo phenomena to become apparent. Quantum field theory requires an array of Bloch Cubes with particle existence represented by a $|z\rangle$ state and its absence represented by $|-z\rangle$ state.

## COMPARISON WITH TRADITIONAL INSTRUCTION

The approach to teaching quantum mechanics described here carefully postpones some topics. Specifically, the TDSE is not explicitly solved, and the unitary operators that represent time evolution are not derived from the TDSE. Hamiltonians, which appear prominently during traditional instruction, do not appear anywhere. The main reason for this choice is that calculus, and the more complex mathematics surrounding exponentiation of operators and coupled ordinary differential equations, is generally too advanced for introductory students. However, the concept of eigenstates is introduced to students. In terms of the Bloch Cube, the eigenstates of the rotation operator are the two states that lie on the axis about which (unitary) rotation takes place. In the approach described here, unitary evolution takes intellectual precedence over the Hamiltonians that generate them. From a pedagogical perspective, it is fairly standard to introduce momentum conservation for a free particle and then later on "derive" it by showing that the momentum operator is the generator of translation in space. Similar pedagogical choices are made in introductory physics, where the forces involved in completely inelastic collsions are "integrated out", yielding a momentum conservation law that does not depend on the details of the collision. This approach to quantum mechanics is not a substitute for traditional approaches to teaching quantum mechanics–it is instead an introduction that can empower students to understand core quantum concepts years before they learn them in a standard physics sequence.

## CONCLUSION

Here we introduced an approach to teaching quantum mechanics to students who have some exposure to vectors and only limited experience with physics. Dirac braket notation is introduced through a close analogy with unit vectors, requiring one extension that can be used to create dot-vec operators. The Bloch cube simplifies the number of quantum states, and eliminates the need for calculus. A number of exercises, more than are described here, can be paired with Bloch cubes, and calculations can be verified and made intuitive through manipulation of the Bloch cube. The knowledge and intuition derived from this approach can support students in more advanced introductory quantum courses, and can help democratize quantum science and engineering by broadening its appeal at an earlier stage of intellectual development.

Following are example comments from first-year college students on instructor evaluation:

- "I loved how we learned about different things using the cube. The class didn't just give me information, it also taught me how to think about things differently and accept the weirdness of quantum ideas."

- "Excellent introduction to quantum physics at an introductory level."

- "The cubes are good."

- "Material presented was not intimidating."

- "I loved how [instructor] helped us learn using cubes and I loved the topic and I found it deeply interesting."

JL gratefully acknowledges funding support from NSF DMR-2225888. CS gratefully acknowledges funding support from NSF PHY-2309260. We thank Robert Devaty for helpful feedback in writing the manuscript.

## AUTHOR DECLARATIONS

### Conflict of Interest

All authors declare that they have no conflicts of interest.

---

# Supplemental Materials
# Teaching Quantum Formalism and Postulates to First-Year Undergraduates

## I. NOTES FOR INSTRUCTORS

Here we list some specific exercises that can be used to introduce quantum mechanics to first-year undergraduate students.

### A. Vectors and unit vectors in two dimensions

A general vector $\vec{V}$ can be expressed as $\vec{V} = V_x\hat{x} + V_y\hat{y}$. Suppose we have two vectors $\vec{V_1}$ and $\vec{V_2}$. We can write $\vec{V_1} = V_{1x}\hat{x} + V_{1y}\hat{y}$ and $\vec{V_2} = V_{2x}\hat{x} + V_{2y}\hat{y}$. If $\vec{V_1} = \vec{V_2}$, students can be asked to show that $V_{1x} = V_{2x}$ and $V_{1y} = V_{2y}$. Similar relations can be inferred if $\vec{V_1} = -\vec{V_2}$ or $\vec{V_1} = 3\vec{V_2}$. Other reviews of vectors, unit vectors, and their properties would be appropriate here.

### B. Dot products and introduction to "Dot-Vec" notation

A general vector $\vec{V}$ can be written as:

$$\vec{V} = V_x\hat{x} + V_y\hat{y} \qquad (1)$$

where $V_x = \hat{x} \cdot \vec{V}$ and $V_y = \hat{y} \cdot \vec{V}$. Unit vectors obey the standard set of orthogonality relations: $\hat{x} \cdot \hat{x} = 1$, $\hat{x} \cdot \hat{y} = 0$, $\hat{y} \cdot \hat{x} = 0$, $\hat{y} \cdot \hat{y} = 1$.

A similar expression can be written for a vector $\vec{U} = U_x\hat{x} + U_y\hat{y}$. As an exercise, students can calculate $\vec{U} \cdot \vec{V}$ by means of the orthogonality relations for $\hat{x}$ and $\hat{y}$. This exercise can be conducted with the abstract expressions for $\vec{U}$ and $\vec{V}$, or with specific values of the components for the vectors.

To help make a connection with Dirac notation, we define a new type of "dot-vector" $\vec{W}\cdot$ with the property $\vec{W}\cdot$ when left-multiplied by $\vec{V}$ yields $\vec{W} \cdot \vec{V}$, the scalar product defined previously.

As an exercise, students can show that if $\vec{W_1}\cdot = \vec{W_2}\cdot$, where $\vec{W_1}\cdot = W_{1x}\hat{x}\cdot + W_{1y}\hat{y}\cdot$ and $\vec{W_2}\cdot = W_{2x}\hat{x}\cdot + W_{2y}\hat{y}\cdot$, then $W_{1x} = W_{2x}$ and $W_{1y} = W_{2y}$. Similar calculations demonstrating how linear combinations of dot-vectors can be constructed using the same rules that apply to vectors, will help students become comfortable with dot-vectors.

### C. More examples of Dot-Vec notation and coordinate transformations

Objects which have the form $\hat{u}\hat{v}\cdot$, where $\hat{u}$ is a unit vector, and $\hat{v}\cdot$ is a unit dot-vector, transform vectors into other vectors. These objects are referred to as "operators" because of the way they operate on vectors to produce other vectors. It is important for students to understand that the operators act linearly. The action of the sum of operator terms can be decomposed and calculated individually.

An example students can explore is defined by $\boldsymbol{R}(\theta) = \cos\theta\hat{x}\hat{x}\cdot - \sin\theta\hat{x}\hat{y}\cdot + \sin\theta\hat{y}\hat{x}\cdot + \cos\theta\hat{y}\hat{y}\cdot$ which corresponds to an operator that rotates vectors by angle $\theta$ about the origin. This exercise can be made more concrete by writing $\boldsymbol{R}(\theta)$ for specific values of $\theta$, e.g., $90°$. Students can use graph paper to create a series of vectors, and explore how this operator transforms the vectors. They can be asked if these operators conform to their intuition about how objects rotate in the $xy$ plane.

A second example of an operator $\boldsymbol{F} = \hat{x}\hat{y}\cdot + \hat{y}\hat{x}\cdot$ simulates the action of flipping over a coin. If $\hat{x}$ represents the heads state of a coin and $\hat{y}$ represents the tails state of a coin, then $\boldsymbol{F}$ acts to flip the state from heads to tails and vice-versa. Students can calculate $\boldsymbol{F}^2$ to show that it equals the identity operator $\boldsymbol{1} = \hat{x}\hat{x}\cdot + \hat{y}\hat{y}\cdot$, corresponding to two flips of the coin which does not change the state.

### D. Dirac notation and analogy with Dot-Vec notation

There is a one-to-one analogy between Dot-Vec notation and Dirac notation. To help reinforce this relationship, students can be asked to fill in the missing parts of Table I:

TABLE I. Dot-Vec / Bra-Ket notation analogy. Fill in the blank fields.

| Dot-Vec | Bra-Ket |
| --- | --- |
| $\hat{x}$ | $|+z\rangle$ |
| $\hat{y}$ | $|-z\rangle$ |
| $\hat{x} \cdot \hat{x} = 1$ | |
| | $\langle +z|-z\rangle = 0$ |
| $\hat{y} \cdot \hat{x} = 0$ | |
| | $\langle -z|-z\rangle = 1$ |
| $\hat{x}\hat{x}\cdot$ | |
| | $|z\rangle\langle -z|$ |
| $\hat{y}\hat{x}\cdot$ | |
| | $|-z\rangle\langle -z|$ |

Another exercise can ask students to identify what is wrong with various expressions, e.g., $|w\rangle = a|+z\rangle + b\langle -z|$, or $D = \langle +z|+z\rangle + |-z\rangle\langle -z|$.

### E. Bloch Cube and representation of quantum states

After handing out a Bloch Cube to every student (they can be made from blank acrylic dice and permanent markers), the upper face and orientation of a Bloch Cube can be identified with 24 states (which are not distinct), given by $\{1, i, -1, -i\} \otimes \{|\pm x\rangle, |\pm y\rangle, |\pm z\rangle\}$. Opposite faces of the cube represent distinct pairs of states, e.g., $|+z\rangle$ and $|-z\rangle$ or $|+x\rangle$ and $|-x\rangle$. Students can be asked whether $|z\rangle$ and $|x\rangle$ are distinct states (they are not). A set of useful formulae can be given to represent the states in the Z basis:

$$|+x\rangle = \frac{1}{\sqrt{2}}(|+z\rangle + |-z\rangle) \tag{2a}$$

$$|-x\rangle = \frac{1}{\sqrt{2}}(|+z\rangle - |-z\rangle) \tag{2b}$$

$$|+y\rangle = \frac{1}{\sqrt{2}}(|+z\rangle + i|-z\rangle) \tag{2c}$$

$$|-y\rangle = \frac{1}{\sqrt{2}}(|+z\rangle - i|-z\rangle) \tag{2d}$$

It can be pointed out to students that $|+x\rangle$ has "some" of $|+z\rangle$ in it, and therefore is not distinct from $|+z\rangle$. States of the Bloch Cube that are joined by an edge are in general not distinct. Only opposite faces are distinct.

### F. Bloch Cube and Born rule

The Born rule determines the probabilistic outcome of measurement. With Bloch Cubes, there are only three types of questions ("X Question", "Y Question", "Z Question"). First we consider the Z Question. Students take their Bloch Cubes and roll them to a random initial state. We can label that initial state the $|w\rangle$ state. Asking the Z Question causes the state to "collapse" into the $|+z\rangle$ state with probability $|\langle w|+z\rangle|^2$, and into into the $|-z\rangle$ state with probability $|\langle w|-z\rangle|^2$. Students can verify that $|\langle w|-z\rangle|^2 + |\langle w|-z\rangle|^2 = 1$. With these formulae, students can calculate that if $w = z$, the outcome of a measurement of Z is deterministic. If $w \neq z$, then the probability of collapsing into $|z\rangle$ and $|-z\rangle$ are both 50%.

### G. Bloch Cube and quantum dynamics

Quantum evolution is governed by Unitary operators that take the form $\boldsymbol{U} = |+v\rangle\langle +u| + |-v\rangle\langle -u|$. The act of transforming a state via Unitary evolution can be illustrated with the Bloch Cube. Rotation of the Bloch Cube about the Z axis does not transform the $|+z\rangle$ or $|-z\rangle$ states, except for a phase. However, it does transform the $|\pm x\rangle$ and $|\pm y\rangle$ states. Students can demonstrate these properties for themselves with their Bloch Cubes.

### H. Proving that unitary operators are represented by Bloch Cube rotation

These transformations can be visualized by spinning the Bloch Cube about the Z axis by a quarter turn and noticing how the sides of the cube transform. The operator corresponding to this operation can be represented as: $\boldsymbol{Z} = |+z\rangle\langle +z| + i|-z\rangle\langle -z|$, and students can be asked to show that $\boldsymbol{Z'} = |+y\rangle\langle +x| + |-y\rangle\langle -x|$ is a valid re-expression (i.e., $\boldsymbol{Z'} = \boldsymbol{Z}$), using the definitions in Eq. 2.

## II. SOME ASSESSMENT QUESTIONS

Useful Formulae:

$$|+x\rangle = \frac{1}{\sqrt{2}}(|+z\rangle + |-z\rangle)$$

$$|-x\rangle = \frac{1}{\sqrt{2}}(|+z\rangle - |-z\rangle)$$

$$|+y\rangle = \frac{1}{\sqrt{2}}(|+z\rangle + i|-z\rangle)$$

$$|-y\rangle = \frac{1}{\sqrt{2}}(|+z\rangle - i|-z\rangle)$$

1. Answering the "Z question". Take one of your Bloch cubes and roll it. For the state that is facing up (call it the "w" state), calculate the probability that for the $|w\rangle$ state you will collapse into the $|+z\rangle$ and $|-z\rangle$ state after asking the Z question. You should do this with help from the Bloch cube, and also by rewriting the $|w\rangle$ state in the Z basis. (e.g., $|+x\rangle = (|+z\rangle + |-z\rangle)/\sqrt{2}$. The probability of collapsing into $|+z\rangle$ is given by $|\langle w|+z\rangle|^2$ and the probability of collapsing into $|-z\rangle$ is given by $|\langle w|-z\rangle|^2$.

2. Roll two Bloch cubes (call them $|w\rangle$ and $|v\rangle$), and write the state $|w\rangle$ in the V (= X, Y, or Z) basis. For example, to write $|-y\rangle$ in the X basis, you multiply by $\hat{1} = |+x\rangle\langle +x| + |-x\rangle\langle -x|$. That is, $|-y\rangle = (|+x\rangle\langle +x| + |-x\rangle\langle -x|)|-y\rangle = (\langle x|-y\rangle|+x\rangle + \langle -x|-y\rangle|-x\rangle$. To evaluate $\langle +x|-y\rangle$, you can write it in a common basis like the Z basis, using the Useful Formula above: $\langle +x|-y\rangle = (\langle +z| + \langle -z|)(|+z\rangle + i|-z\rangle)) = (1+i)/2$.

3. We showed in class that $\hat{R} = |+y\rangle\langle +x| + |-y\rangle\langle -x|$ can be rewritten as $\hat{R}' = |+z\rangle\langle +z| + i|-z\rangle\langle -z|$. Show explicitly that $\hat{R}|z\rangle = |z\rangle$ by writing everything in the Z basis.